\documentstyle[aps,amstex,amssymb,preprint,epsf]{revtex}
\tightenlines
\begin{document}                                                       

\draft 

                                            
\title {Euclidean resonance and a new type of nuclear reactions}

\author{Boris Ivlev} 

\address
{Department of Physics and Astronomy\\
University of South Carolina, Columbia, SC 29208\\
and\\
Instituto de F\'{\i}sica, Universidad Aut\'onoma de San Luis Potos\'{\i}\\
San Luis Potos\'{\i}, S. L. P. 78000 Mexico}

\maketitle

\begin{abstract}
The extremely small probability of quantum tunneling through an almost classical potential barrier may become not small
under the action of the specially adapted nonstationary field. The tunneling rate has a sharp peak as a function of the
particle energy when it is close to the certain resonant value defined by the nonstationary field (Euclidean resonance). 
Alpha decay of nuclei has a small probability since the alpha particle should tunnel through a very nontransparent nuclear
Coulomb barrier. The incident proton, due to the Coulomb interaction with the tunneling alpha particle, plays the role of 
a nonstationary field which may result in Euclidean resonance in tunneling of the alpha particle. At the resonant proton
energy, which is of the order of $0.2$\hspace{0.1cm}MeV, the alpha particle escapes the nucleus and goes to infinity with 
no influence of the nuclear Coulomb barrier. The process is inelastic since the alpha particle releases energy and the 
proton gains it. This stimulation of alpha decay by a proton constitutes a new type of nuclear reaction. 

\end{abstract} \vskip 1.0cm
   

\narrowtext
\section{INTRODUCTION}
\label{sec:intro}
A control of processes of quantum tunneling through potential barriers by external signals is a part of the field called 
quantum control which is actively developed now, see, for example, Ref.~\cite{RABITZ} and references therein. Excitation of
molecules, when one should excite only particular chemical bonds \cite{SHI,JUDSON,KOHLER}, formation of programmable atomic
wave packets \cite{SCHUMACHER}, a control of electron states in heterostructures \cite{KRAUSE}, and a control of photocurrent
in semiconductors \cite{ATANASOV}, are typical examples of control by laser pulses. A control of quantum tunneling through 
potential barriers is also a matter of interest, since tunneling is a part of many processes in nature. The computation of 
probability for a classically forbidden region has a certain peculiarity from the mathematical stand point: there
necessarily arises here the concept of motion in imaginary time or along a complex trajectory \cite{LANDAU,POKR,COLEMAN}.
The famous semiclassical approach of Wentzel, Kramers, and Brillouin (WKB) \cite{LANDAU} for tunneling probability can be
easily reformulated in terms of classical trajectories in complex time. The method of complex trajectories is also
applicable to a nonstationary case \cite{KELDYSH,PERELOMOV}. The method has been further developed in papers 
\cite{MELN1,MELN2,MELN3,MELN4,MELN5}, where singularities of the trajectories in the complex plane were accounted for an 
arbitrary potential barrier (see also \cite{MILLER}). Recent achievements in the semiclassical theory are presented in 
Refs.\cite{KESHA,DEFENDI,MAITRA,ANKERHOLD,CUNIBERTI}.

Let us focus on the main aspects of tunneling under nonstationary conditions. When the electric field ${\cal E}\cos\Omega t$
acts on a tunneling particle of the initial energy $E$, it can absorb the quantum $\hbar\Omega$ (with the probability
proportional to the small parameter ${\cal E}^{2}$) and tunnel after that in a more transparent part of the barrier with
the higher energy $E+\hbar\Omega$. The pay in the absorption probability may be compensated by the probability gain in 
tunneling. In this case the system tends to absorb further quanta to increase the total probability of passing the barrier.
This mechanism of barrier penetration is called photon-assisted tunneling. If $\hbar\Omega$ is not big, the process of 
tunneling, with the simultaneous multiquanta absorption, can be described in a semiclassical way by the method of 
classical trajectories in the complex time \cite{MELN1,MELN2,MELN3,MELN4,MELN5}. When a tunneling particle of the energy 
$E$ is acted by a short-time pulse, the tunneling probability is associated with the particle density carrying away in
the outgoing wave packet. The particle energy after escape is $E+\delta E$, where the energy gain $\delta E=N\hbar\omega$, should
be extremized with respect to the number of absorbed quanta $N$ and the energy $\hbar\omega$ of each quantum \cite{IVLEV1}. 

According to the perturbation theory, the both probabilities, of absorption and emission of quantum, are small being
proportional to ${\cal E}^{2}$. After absorption, $\delta E$ is positive which enhances the total probability due to increase
of the tunneling rate. After emission, $\delta E$ is negative and the particle should tunnel with a smaller energy in a
less transparent part of the barrier; in this case there is no gain in the probability due to tunneling as for absorption.
At first sight, tunneling in a nonstationary field cannot be assisted by an emission of quanta of this field since one 
should pay in probability twice. Thus, one can expect the double loss in probability due to the emission of quanta and the 
reduction of tunneling transparency. This conclusion is correct as soon as the nonstationary field is small and the 
perturbation theory is applicable. 

Under increase of the nonstationary field the process of tunneling with emission of quanta becomes completely different 
compared to one, predicted on the basis of the perturbation theory \cite{IVLEV2}. The crucial role here plays the fact, 
that the non-perturbative wave function is mainly determined by its big phase (real or imaginary). In this case there is 
no the double loss in probability. The quantum process is not simply reduced to separate emission and tunneling. However,
the phase behavior can be interpreted in the way of enhancement of the total probability due to emission processes which
competes now with the reduction of the tunneling transparency. The competition between the enhancement of the total 
probability due to emission and the reduction of it due to tunneling results in the unexpected effect: the total 
probability (defined as a particle density, carrying away by the outgoing wave packet) becomes not exponentially small for
the certain particle energy $E_{R}$ in the well. This energy depends on parameters of the nonstationary field and in a
vicinity of $E_{R}$ the probability sharply peaks as a function of energy. This reminds, formally, a resonant behavior and
is called Euclidean resonance \cite{IVLEV2}. The energy $E_{R}$ can be called the resonant energy. In \ref{sec:eucl} this 
phenomenon is deduced from the analysis of the quantum mechanical phases using only simple physical arguments. 

As well known, nuclear Coulomb barriers may be very significant in nuclear physics, playing a role of blockade for particle
approach or escape \cite{WEISSKOPF,BOHR}. The famous example of such a nuclear process is alpha decay of nuclei which has
a small probability since the alpha particle should penetrate through a very non-transparent nuclear Coulomb barrier 
\cite{WEISSKOPF,BOHR}. As any tunneling process, alpha decay can be influenced by a nonstationary field. The duration of 
artificially generated pulses (see, for example, \cite{EFIMOV}) is too long compared to the nucleus characteristic time of
$10^{-21}$\hspace{0.1cm}s which makes impossible their influence on alpha decay. A role of a nonstationary field can be 
played by a moving charged particle which collides the decaying nucleus. For example, a proton of the energy of 
$1$\hspace{0.1cm}Mev sweeps the tunneling region in, approximately, $10^{-21}$\hspace{0.1cm}s. An incident proton, moving 
towards a tunneling alpha particle, reduces its energy due to the Coulomb interaction between them. Since the energy is 
lost but not gained during tunneling, Euclidean resonance may be expected. As shown below, this happens. At the certain 
energy of the incident proton the simultaneous tunneling of alpha particle has not an exponentially small probability, 
according to Euclidean resonance.
 
In other words, due to interaction with an incident proton of the certain energy, the alpha particle escapes the 
nucleus and goes to infinity with no influence of the Coulomb barrier. This stimulation of alpha decay by a proton
constitutes the new type of nuclear reaction. Normally, nuclear reactions involve strong interaction at nuclear distances.
In this case, only the short range start of the alpha particle from the nucleus is due to strong interaction and the 
main physics occurs further, at the larger distance where strong forces do not act and only Coulomb effects are involved. 
This new type of nuclear reactions has a very resonant character with respect to an energy of the incident proton. As 
shown below, the typical resonance energy of a proton is in the range of $0.2$\hspace{0.1cm}MeV. One should emphasize, that 
this resonance results from solely Coulomb effects in contrast, for example, to resonances in nuclear physics due to 
formation of compound nuclei.  
\section{PHOTON-ASSISTED TUNNELING}
\label{sec:ph-as}
A penetration of a particle through a potential barrier is forbidden in classical mechanics. Only due to quantum effects 
the probability of passing across a barrier becomes finite and it can be calculated on the basis of WKB approach, which is 
also called the semiclassical theory. The transition probability through the barrier, shown in Fig.~\ref{fig1}, is
\begin{equation} 
\label{1}
W\sim\exp\left[-A_{0}(E)\right]
\end{equation}
where
\begin{equation}
\label{2}
A_{0}(E)=\frac{2}{\hbar}\int dx\sqrt{2m\left[V(x)-E\right]}
\end{equation}
is the classical action measured in units of $\hbar$. The integration goes under the barrier between two classical turning 
points where $V(x)=E$. One can use the general estimate $A_{0}\sim V/\hbar\omega$, where $V$ is the barrier height and $\omega$ is
the frequency of classical oscillations in the potential well. A semiclassical barrier relates to a big value 
$V/\hbar\omega\gg 1$. 

What happens when the static potential barrier $V(x)$ is acted by a weak nonstationary electric field ${\cal E}(t)$? In this
case there are two possibilities for barrier penetration: (i) the conventional tunneling, which is not affected by 
${\cal E}(t)$, shown by the dashed line in Fig.~\ref{fig2}(a), and (ii) an absorption of the quantum $\hbar\Omega$ of the field
${\cal E}(t)$ and subsequent tunneling with the new energy $E+\hbar\Omega$. The latter process is called photon-assisted 
tunneling. The total probability of penetration across the barrier can be schematically written as a sum of two 
probabilities
\begin{equation}
\label{3}
W\sim\exp\left(-\frac{V}{\hbar\omega}\right)+
\left(\frac{a\hspace{0.05cm}{\cal E}_{\Omega}}{\hbar}\right)^{2}\exp\left(-\frac{V-\hbar\Omega}{\hbar\omega}\right)
\end{equation}
where ${\cal E}_{\Omega}$ is the Fourier component of the field ${\cal E}(t)$. The second term in Eq.~(\ref{3}) relates to 
photon-assisted tunneling and it is a product of probabilities of two quantum mechanical processes: absorption of the 
quantum $\hbar\Omega$ and tunneling through the reduced barrier $V-\hbar\Omega$. The length $a$ is a typical barrier extension 
in space. When the frequency is high $\Omega >\omega$, the second term starts to dominate at sufficiently small nonstationary
field $(a{\cal E}_{\Omega}/\hbar)^{2}>\exp(-\Omega/\omega)$. This is a feature of tunneling processes. Normally a nonstationary field
starts to dominate at bigger amplitudes $(a{\cal E}_{\Omega}/\hbar)^{2}>1$. When the second term in Eq.~(\ref{3}) exceeds the 
first one, further orders of perturbation theory should be accounted which correspond to the multiple absorption, shown in 
Fig.~\ref{2}(b). 

Let us specify a shape of a field pulse in the form
\begin{equation}
\label{4}
{\cal E}(t)=\frac{{\cal E}}{1+t^{2}/\theta^{2}}
\end{equation}
with the Fourier component ${\cal E}_{\Omega}=\pi{\cal E}\theta\exp(-\theta|\Omega|)$. In this case, in addition to the steady 
particle flux from the barrier, an outgoing wave packet is created which carries away the certain particle density. Then 
the probability $W$ of the transition through the process of absorption of $N$ quanta and subsequent tunneling with the 
higher energy $E+N\hbar\Omega$, shown in Fig.~\ref{2}(b), is
\begin{equation} 
\label{5}
W\sim\left(\frac{a{\cal E}_{\Omega}}{\hbar}\right)^{2N}\exp\left[-A_{0}(E+N\hbar\Omega)\right]=
\left(\frac{\pi}{\hbar}\hspace{0.05cm}\theta a{\cal E}\right)^{2N}\exp(-A)
\end{equation}
where
\begin{equation}
\label{6}
A=\frac{2\theta}{\hbar}\hspace{0.1cm}\delta E + A_{0}(E+\delta E)
\end{equation}
Here the total energy transfer $\delta E = N\hbar\Omega$ is introduced ($\Omega > 0$). The maximum transition probability 
through the barrier is determined by some finite value of $\delta E$, which provides a minimum of $A$ and is defined by the
condition $\partial A(E+\delta E)/\partial\delta E = 0$. An existence of such a minimum is possible if a grow up of small $\delta E$ 
reduces $A$, that is, under the condition $2\theta < \hbar\mid\partial A_{0}(E)\partial E\mid$ of sufficiently short pulses.
In other words, sufficiently short and not very small pulses (however, still much smaller than the static barrier field)
strongly enhance tunneling by photon assistance.

Eq.~(\ref{5}) omits some details and it is rather an illustration of a tunneling mechanism with quanta absorption. For
example, the accurate perturbation theory starts with the linear ${\cal E}$-term. The exact nonperturbative theory, which 
is a generalization of the conventional semiclassical approach, results in the same $A$. The semiclassical approach is 
also sensitive to the sign of ${\cal E}$ since the classical energy transfer is determined by 
${\cal E}(t)\hspace{0.05cm}\partial x/\partial t$, where $x(t)$ is a classical particle trajectory. The strong photon-assisted 
tunneling exists only at positive ${\cal E}$.
\section{EUCLIDEAN RESONANCE}
\label{sec:eucl}
Besides the absorption of quanta in Fig.~\ref{fig2}, also the emission is possible, shown in Fig.~\ref{fig3}. This process is
provided by negative frequencies $\Omega <0$ and the energy transfer is also negative $\delta E <0$. In this case Eq.~(\ref{5})
gives $A=2\theta|\delta E|/\hbar +A_{0}(E-|\delta E|)$ which does not correspond to any extreme since 
$\partial A/\partial|\delta E|\hspace{0.1cm}>0$. So, on the basis of perturbation theory, one can conclude that emission
processes cannot enhance a barrier penetration at least for the pulse amplitude on the border of applicability of the
perturbation theory $a{\cal E}\lesssim\hbar/\theta$. May an emission process enhance a tunneling rate when the signal 
amplitude is not small $\hbar/\theta <a{\cal E}$?
\subsection{Phase connection}
At the big pulse amplitude ${\cal E}$ its contribution to the classical action, generally speaking, is big, compared to 
Planck's constant,
\begin{equation}
\label{7}
a{\cal E}\theta\gg\hbar
\end{equation}
and the wave function $\psi\sim\exp(i\chi)$ is mainly determined by its big phase $\chi$, which can be imaginary as well. 
Another condition of a big phase is a slow varying pulse
\begin{equation}
\label{8}
\frac{\hbar}{V}\ll\theta
\end{equation}
since a big phase should be built up during a long time. Here $V$ is the barrier height and, therefore, $\hbar/V$ is some
intrinsic time of the problem. One has to note, that within the conditions (\ref{7}) and (\ref{8}) the pulse amplitude can
be still less compared to the static field of the barrier $V/a$.

Suppose $x_{1}$ and $x_{2}$ to relate to the classical turning points (1) and (2) in Fig.~\ref{fig3}. At 
$t\rightarrow\pm\infty$, when ${\cal E}(t)=0$, there is a conventional tunneling through the barrier. For a symmetric in time 
pulse ${\cal E}(t)$ the modulus of the wave function of the outgoing particle ${|\psi(x_{2},t)|}$ has an extreme (maximum) 
value at $t=0$ which relates to the maximum amplitude of the outgoing wave packet at $t>0$. According to Feynman 
\cite{FEYNMAN}, an extreme wave function corresponds to a classical trajectory of the particle connecting the points 
$\{x_{1},0\}$ and  $\{x_{2},0\}$. But in the present situation there are no classical trajectories under a barrier.

However, one can find a connection between two constants $\psi(x_{1},0)$ and $\psi(x_{2},0)$ without an exact solution of
Schr{\"o}dinger equation. The main point of this procedure is the possibility to define the wave function mainly by its 
big phase (real or imaginary) which is true under the conditions (\ref{7}) and  (\ref{8}). Then one can consider a particle 
state at any coordinate as in classical mechanics. Since the moment $t=0$ corresponds to the extreme situation of maximum 
output, one can look for an alternative extreme way to connect phases of $\psi(x_{1},0)$ and $\psi(x_{2},0)$. Let us find some 
formal path from (1) to (2) which relates to an extreme total phase with respect to the energy of the final state (2).    
  
Suppose, that the particle state $(i)$ at $x=x_{i}$ is chosen with the same energy $E-|\delta E|$ as at the point (2) in 
Fig.~\ref{fig4}. The particle from the position $(i)$ can tunnel to the position (2) and the acquired phase
is imaginary
\begin{equation}
\label{9}
\psi(x_{2},0)\sim\psi(x_{i},0)\exp\left[-\frac{1}{2}A_{0}(E-|\delta E|)\right]
\end{equation}
On the other hand, the particle can absorb $N=|\delta E|/\hbar\Omega$ quanta, as in Fig.~\ref{fig4}(a), and go to the state (1) 
where $\psi(x_{1},0)\sim\psi(x_{i},0)\left({\cal E}_{\Omega}\right)^{N}$. This also leads to an imaginary acquired phase 
\begin{equation}
\label{10}
\psi(x_{1},0)\sim\psi(x_{i},0)\exp\left(-\frac{\theta}{\hbar}|\delta E|\right)
\end{equation}
We use here the expression for ${\cal E}_{\Omega}$. Analogously, the particle can emit $N$ quanta and go from (1) to $(i)$, as 
in Fig.~\ref{fig4}(b). This leads to another connection of (1) and $(i)$ 
\begin{equation}
\label{10a}
\psi(x_{i},0)\sim\psi(x_{1},0)\exp\left(-\frac{\theta}{\hbar}|\delta E|\right)
\end{equation}
According to Eqs.~(\ref{9}) - (\ref{10a}), the connection between (1) and (2) can be written in the form
\begin{equation}
\label{10b}
\psi(x_{2},0)\sim\exp\left[-\frac{1}{2}A_{0}(E-|\delta E|)\right]
\left[c_{a}\exp\left(\frac{\theta}{\hbar}|\delta E|\right)+ c_{b}\exp\left(-\frac{\theta}{\hbar}|\delta E|\right)\right]\psi(x_{1},0)
\end{equation}
where $c_{a}$ and $c_{b}$ are constants. Only the first term in Eq.~(\ref{10b}) provides an extreme of the total phase and, 
therefore, one should put $c_{a}=1$ and $c_{b}=0$. This is an alternative extreme way of phase connection. The relation 
(\ref{10b}) corresponds to a formal (no direct analogy with the exact solution) path connecting the phases (1) and (2).   
The extreme (at the moment $t=0$) transition probability from (1) to (2) is determined by the found extreme phase
difference in (\ref{10b})
\begin{equation}
\label{11}
W\left(1\rightarrow 2\right)\sim{\bigg|\hspace{0.05cm}\frac{\psi(x_{2},0)}{\psi(x_{1},0)}\hspace{0.05cm}\bigg|}^{2}
\sim\exp\left[-A(|\delta E|)\right]
\end{equation}
where
\begin{equation}
\label{12}
A(|\delta E|)=A_{0}(E-|\delta E|)-\frac{2\theta}{\hbar}|\delta E|
\end{equation}
As also follows from the solution of Schr\"{o}dinger equation, $A(|\delta E|)$ weakly depends on the pulse amplitude ${\cal E}$
under the conditions (\ref{7}) and (\ref{8}). The optimum energy transfer $|\delta E_{0}|$ should be determined from the 
extreme condition $\partial A(|\delta E|)/\partial |\delta E|=0$, which reads
\begin{equation}
\label{13}
\frac{2\theta}{\hbar}=-\frac{\partial A_{0}(E-|\delta E|)}{\partial E}
\end{equation}
and determines the extreme value $A=A(|\delta E_{0}|)$. Again, the physical transition probability at $t=0$, $W\sim\exp(-A)$, 
can be considered either as an extreme in $t$ of the exact solution or as an extreme in $|\delta E|$ in the above phase 
connection procedure.

The phase connection, illustrated in Fig.~\ref{fig4} is applicable for big pulses $\hbar/\theta <a{\cal E}$ when the wave 
function is mainly defined by its big phase (real or imaginary) and the phase difference between (1) and $(i)$ is of the 
opposite sign compared to one between $(i)$ and (1). This phase connection, as follows from the solution of Schr\"{o}dinger 
equation, is applicable in our case of decay of a metastable state. But it does not mean, that its applicability is 
automatically valid in other situations with a big phase. For example, a penetration of an incident particle into the 
potential well is not described by this method. 
\subsection{Resonance conditions}
An applicability of the phase connection, besides the conditions (\ref{7}) and  (\ref{8}), is restricted by the inequality
\begin{equation}
\label{14}
\exp(-A)\ll 1
\end{equation}
A pulse amplitude ${\cal E}$ should be negative, since the classical energy transfer is determined through the classical
trajectory $x(t)$  as ${\cal E}(t)\partial x/\partial t$. $A$ is determined by the particle energy $E$ in the well and by the 
pulse duration $\theta$. Suppose $\theta$ to be fixed and the particle energy $E$ to vary. Then the optimum energy transfer
$|\delta E_{0}(E)|$, determined by Eq.~(\ref{13}), is a function of $E$ and one can define the certain energy $E_{R}(\theta)$ as 
a solution of the equation
\begin{equation}
\label{15}
A_{0}\left[E-|\delta E_{0}(E)|\right]-\frac{2\theta}{\hbar}\hspace{0.1cm}|\delta E_{0}(E)|=0
\end{equation}
At the particle energy close to $E_{R}$, $A\simeq 2\theta[E_{R}-E]/\hbar$ and the peak in tunneling probability at $t=0$, related 
to the particle density in the outgoing wave packet, is
\begin{equation}
\label{16}
W\sim\exp\left(-\frac{2\theta}{\hbar}\left[E_{R}(\theta)-E\right]\right)
\end{equation}
The formal applicability of these relations is $W\ll 1$, nevertheless, the probability peak can dramatically grow up
(upon approaching $E_{R}$), for example, from $10^{-37}$ to $10^{-2}$. The energy $E_{R}$ has the order of magnitude of the
barrier height $V$ and, hence, $2\theta(E_{R}-E)/\hbar\sim(\theta V/\hbar)(E_{R}-E)/E_{R}$. This means, that the tunneling probability 
at $t=0$ has a sharp peak, like a resonance, as a function of the particle energy $E$ near $E_{R}$. The effect may be called 
Euclidean resonance when $E_{R}$ plays a role of the resonant energy. The origin of the word ``Euclidean'' is explained 
below.

Euclidean resonance also can be treated in another way. Suppose the energy level in the well to be fixed. Then one can
adjust the pulse parameter $\theta$ to meet the condition $E_{R}=E$ when the barrier becomes almost transparent at $t=0$ for
the energy $E$. Note, that for other particle energies the barrier remains low-transparent.

\begin{figure}[p]
\begin{center}
\vspace{1.5cm}
\leavevmode
\epsfxsize=\hsize
\epsfxsize=10cm
\epsfbox{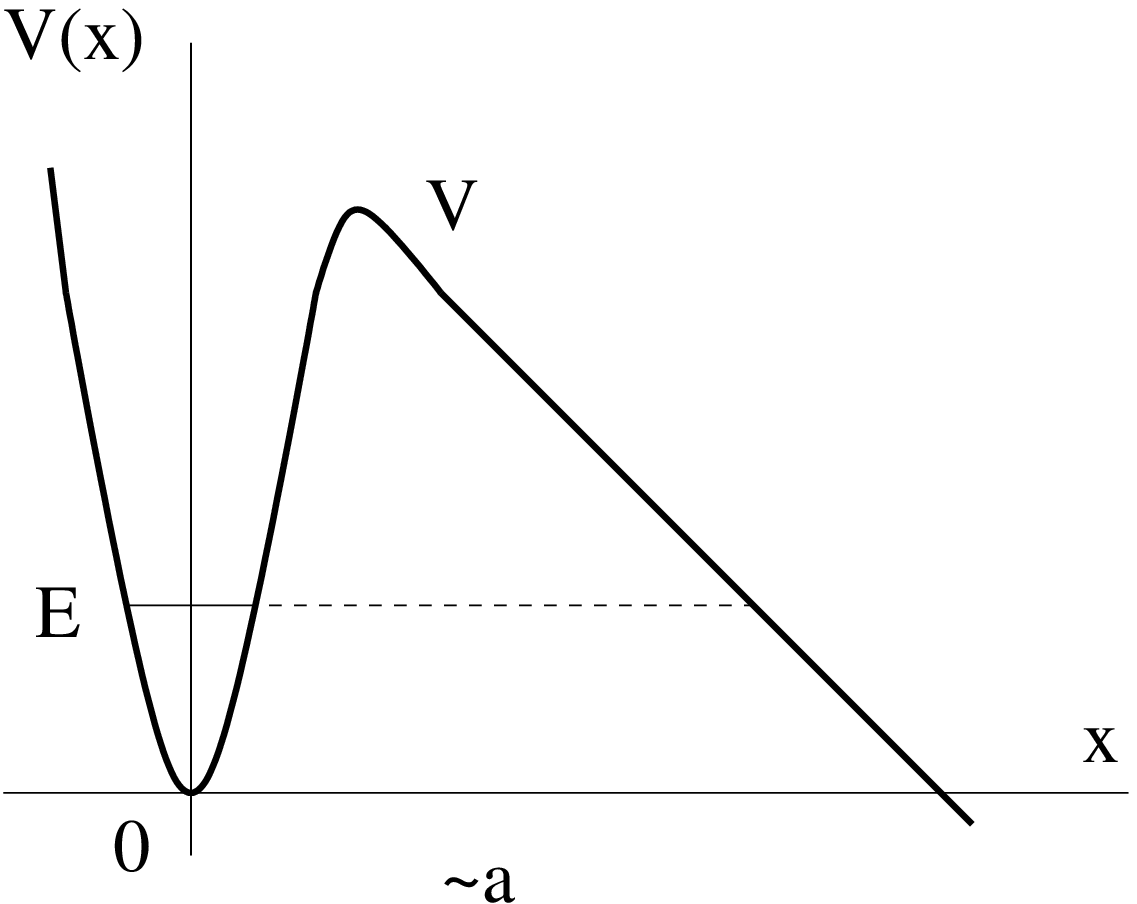}
\vspace{2cm}
\caption{The path of tunneling is denoted by the dashed line. $E$ is the energy of the metastable state, $V$ is the barrier 
height, and $a$ is the typical potential length.}
\label{fig1}
\end{center}
\end{figure}

\begin{figure}[p]
\begin{center}
\vspace{1.5cm}
\leavevmode
\epsfxsize=\hsize
\epsfxsize=12cm
\epsfbox{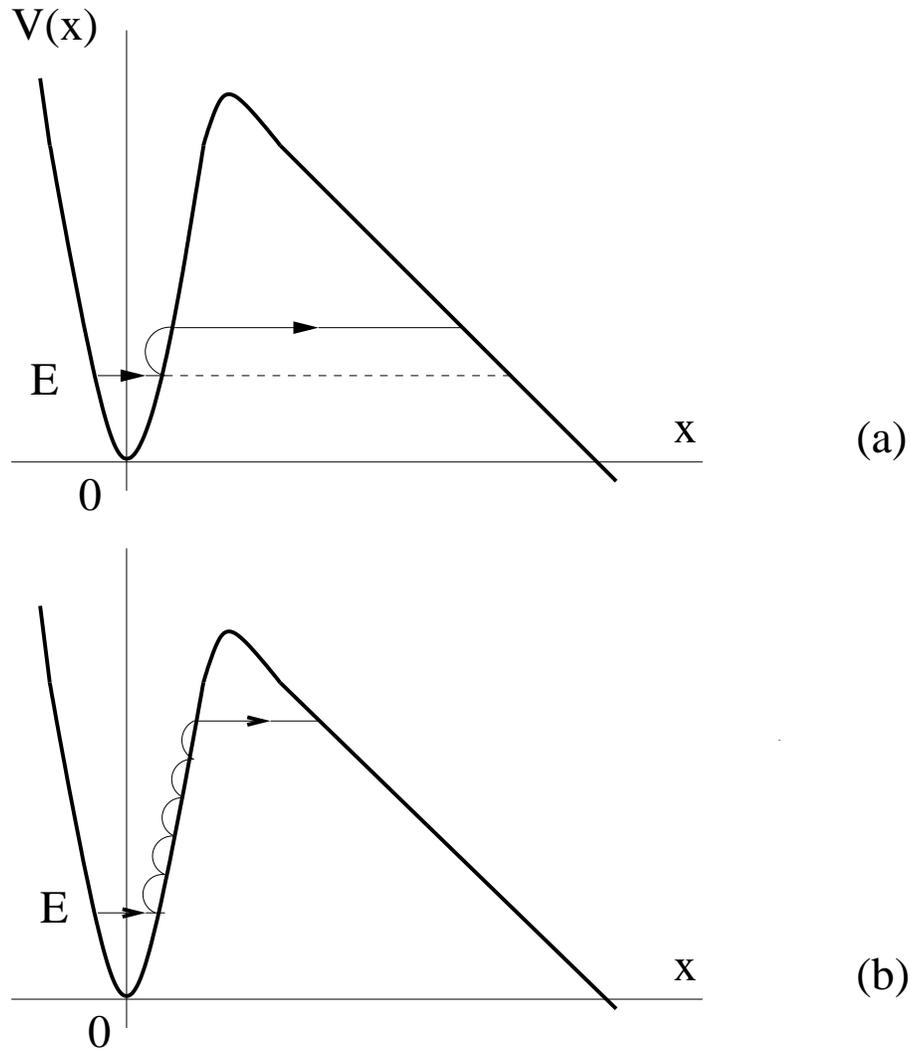}
\vspace{2cm}
\caption{The particle can absorb a quantum and tunnel in a more transparent part of the barrier with the energy 
$E+\hbar\Omega$ $(a)$. The process of the multiquanta absorption with the subsequent tunneling $(b)$.}
\label{fig2}
\end{center}
\end{figure}

\begin{figure}[p]
\begin{center}
\vspace{1.5cm}
\leavevmode
\epsfxsize=\hsize
\epsfxsize=10cm
\epsfbox{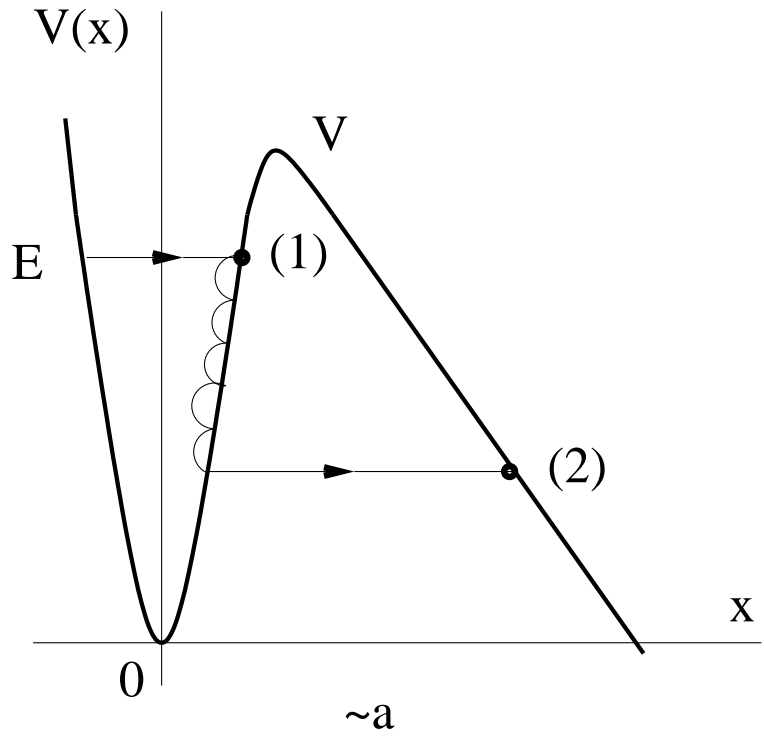}
\vspace{2cm}
\caption{The multiquanta emission with the subsequent tunneling in a less transparent part of the barrier.}
\label{fig3}
\end{center}
\end{figure}

\begin{figure}[p]
\begin{center}
\vspace{1.5cm}
\leavevmode
\epsfbox{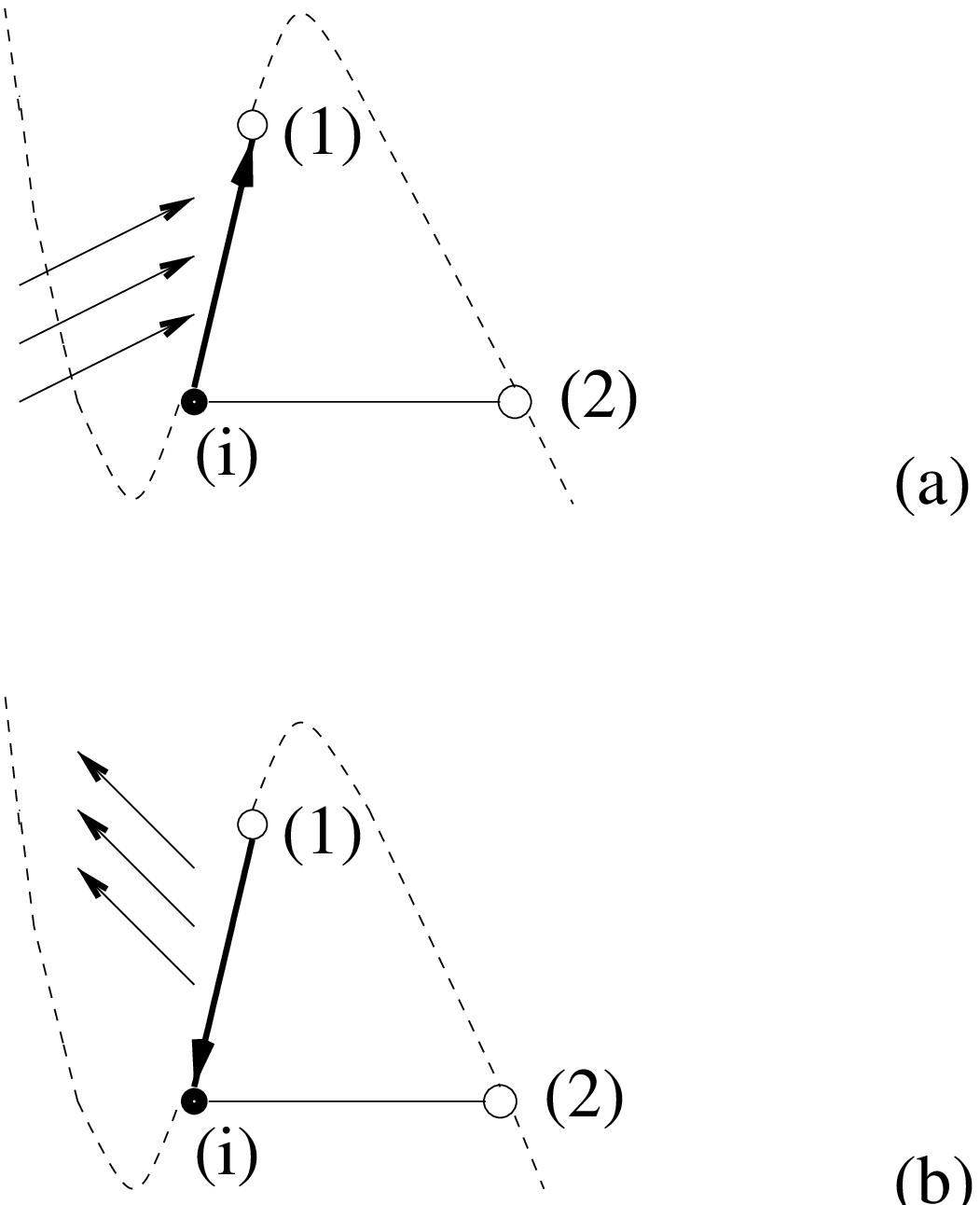}
\vspace{2cm}
\caption{$(1)$ is the initial particle state with the energy $E$ and $(2)$ is the final state of the energy $E-|\delta E|$ 
after tunneling. The state $(i)$ has the same energy as $(2)$. There are two ways to connect phases of the states $(i)$ 
and $(1)$: by quanta absorption $(a)$ and by quanta emission $(b)$.}
\label{fig4}
\end{center}
\end{figure}

In contrast to photon-assisted tunneling, which has a connection with the perturbation theory (see Fig.~\ref{fig2}), the 
phenomenon of Euclidean resonance is completely non-perturbative. One can construct some analogy of Euclidean resonance by
adding the narrow potential well $-v(b)\delta (x-b)$ to the potential $V(x)$ in Fig.~\ref{fig1}. The distance $b$ is within the
under-barrier region and the positive coefficient $v(b)$ is chosen to get the same energy level $E$ in the $\delta$-well
as in the main well (resonance tunneling). Suppose $b$ to be close to the main well in Fig.~\ref{fig1} and the peak of
the wave function at $x=b$ is of the same order as in the main well. Let us move $b$ slowly (compared to the time 
$\hbar/V$) away from the main well to get the big peak of the wave function at some point under the barrier. After this
$v(b)$ is switched off fast and the remaining distribution, which is not exponentially small, goes partly outside the 
barrier. So, this non-stationary mechanism provides the outgoing wave packet which is not exponentially small. As in 
Euclidean resonance, two issues are important in this example: (i) the nonstationary potential should be not small and 
(ii) it should be very precisely chosen to get the condition of resonance tunneling, otherwise, the peak of the wave 
function inside the barrier would be exponentially small. 
\section{AN EXAMPLE OF EUCLIDEAN RESONANCE}
Let us consider an electron emission from a metal, left in Fig.~\ref{fig5}, to the vacuum due to the applied electric field
${\cal E}_{0}+{\cal E}(t)$ where ${\cal E}_{0}$ is a constant. The energy $E$ is supposed not to be above the Fermi level. An
electron emission occurs by tunneling through the barrier $V(x)-x{\cal E}(t)$ where the pulse acts only at $x>0$. The static
barrier is $V(x)=V-x{\cal E}_{0}$ at $x>0$ and $V(x)=0$ at $x<0$. 

The conventional WKB action (\ref{2}) has the form 
\begin{equation}
\label{17}
A_{0}(E)=\frac{4}{3\hbar}\hspace{0.1cm}(V-E)\tau_{00}(E)
\end{equation}
where
\begin{equation}
\label{18}
\tau_{00}(E)=\frac{1}{{\cal E}_{0}}\sqrt{2m(V-E)}
\end{equation}
The relation (\ref{13}) turns to $\tau_{00}(E-|\delta E|)=\theta$ and the optimum energy transfer is 
\begin{equation}
\label{19}
|\delta E_{0}|=\frac{\theta^{2}{\cal E}^{2}_{0}}{2m}-V+E
\end{equation}
Eq.~(\ref{15}) defines the resonance energy
\begin{equation}
\label{20}
E_{R}(\theta)=V-\frac{\theta^{2}{\cal E}^{2}_{0}}{6m}
\end{equation}
One can note, that, under the resonance condition, the input energy $E_{R}$ is connected to the output one $E_{R}-|\delta E|$ 
as $3(V-E_{R})=V-(E_{R}-|\delta E|)$.

These formulae relate to the Lorentzian shape of a pulse (\ref{4}). What happens for the nonstationary field 
${\cal E}(t)={\cal E}\cos\Omega t$ or of some other shape? The above arguments, based on general physical principles, are not 
sufficient to answer this question and a more sophisticated treatment is required. This is considered in 
\ref{sec:traj}.
\section{TRAJECTORIES IN IMAGINARY TIME}
\label{sec:traj}
According to Feynman \cite{FEYNMAN}, when the phase of a wave function is big, it can be expressed through classical 
trajectories of the particle. But in our case there are no conventional trajectories since a classical motion is forbidden
under a barrier. Suppose a classical particle to move in the region to the right of the classical turning point (2) in 
Fig.~\ref{fig5} and to reach the point (2) at $t=0$. Then, close to the point (2), $x(t)=x_{2}+ct^{2}$ ($c>0$) and there is no 
a barrier penetration as at all times $x(t)>x_{2}$. Nevertheless, if $t$ is formally imaginary, $t=i\tau$, the penetration 
becomes possible since $x(i\tau)=x_{2}-c\tau^{2}$ is less then $x_{2}$. Therefore, one can use classical trajectories in 
imaginary time to apply Feynman's method to tunneling.
\subsection{Newton's equation and classical action}
\begin{figure}[p]
\begin{center}
\vspace{1.5cm}
\leavevmode
\epsfxsize=\hsize
\epsfxsize=9cm
\epsfbox{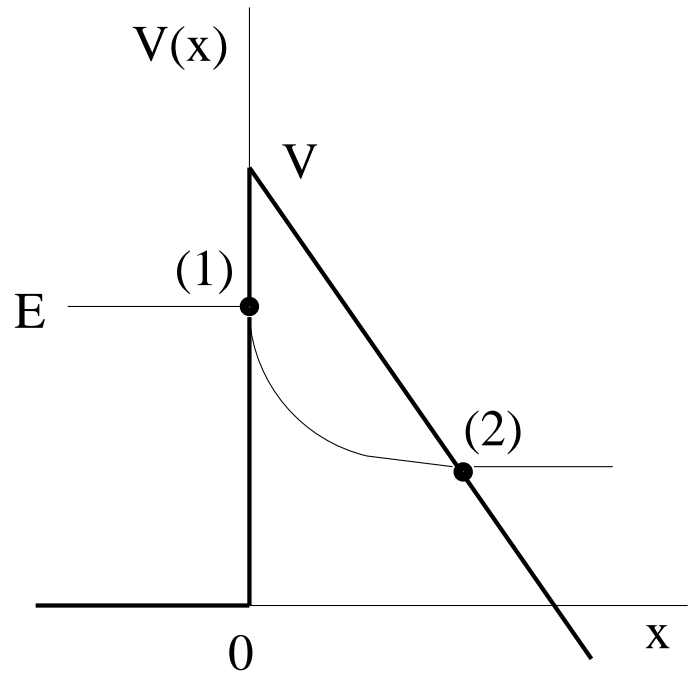}
\vspace{2cm}
\caption{The electron emission from the metal (left) to the vacuum by applying the constant electric field which is weakly
modulated in time.}
\label{fig5}
\end{center}
\end{figure}

A classical trajectory has to satisfy Newton's equation in imaginary time
\begin{equation}
\label{21}
m\hspace{0.1cm}\frac{\partial^{2}x}{\partial\tau^{2}}-\frac{\partial V(x)}{\partial x}=-{\cal E}(i\tau)
\end{equation}
where $V(x)$ is the static barrier in Fig.~\ref{fig5}. The classical turning point (2) in Fig.~\ref{fig5} is reached at
$\tau =0$ with the initial condition
\begin{equation}
\label{22}
\frac{\partial x}{\partial\tau}\hspace{0.1cm}\bigg |_{\tau =0}=0\hspace{1.5cm}\rm{point}\hspace{0.2cm}(2)
\end{equation}
The point (1) is reached at $t=i\tau_{0}$ when
\begin{equation}
\label{23}
x(i\tau_{0})=0\hspace{1.5cm}\rm{point}\hspace{0.2cm}(1)
\end{equation}
The ``time'' $\tau_{0}$ is expressed through the particle energy $E$ before the barrier where the nonstationary field does
not act
\begin{equation}
\label{24}
E=\frac{m}{2}\left(\frac{\partial x}{\partial\tau}\right)^{2}_{\tau_{0}}+V(0)
\end{equation}
The conditions (\ref{22}), (\ref{23}), and (\ref{24}) define the solution $x(i\tau)$ of Eq.~(\ref{21}). This solution, in turn,
defines the extreme (in time) transition probability related to the particle density in the outgoing wave packet
\begin{equation}
\label{25}
W\sim\exp(-A)
\end{equation}
where $A$ is the classical action in units of Planck's constant
\begin{equation}
\label{26}
A=\frac{2}{\hbar}\int^{\tau_{0}}_{0}d\tau\left[\frac{m}{2}\left(\frac{\partial x}{\partial\tau}\right)^{2}+V(x)-x{\cal E}(i\tau)-E\right]
\end{equation}
$\tau_{0}$ can be treated as ``time'' of motion under a barrier. Eq.~(\ref{26}) holds for any barrier which is zero at $x<0$.
Without a nonstationary pulse, ${\cal E}=0$, $\tau_{0}$ coincides with its static value
\begin{equation}
\label{27}
\tau_{00}=\sqrt{\frac{m}{2}}\int\frac{dx}{\sqrt{V(x)-E}}
\end{equation}
and the action (\ref{26}) turns to $A_{0}(E)$ (\ref{2}). The integration in Eq.~(\ref{27}) goes between classical turning 
points where $V(x)=E$. According to classical mechanics,
\begin{equation}
\label{28}
2\tau_{00}=-\frac{\partial A_{0}(E)}{\partial E}
\end{equation}
\subsection{An example of imaginary trajectories}
Let us consider the particular pulse (\ref{4}). In imaginary time ${\cal E}(i\tau)=1/(1-\tau^{2}/\theta^{2})$ diverges at 
$\tau =\theta$ and this sets $\tau_{0}$ in Eq.~(\ref{26}) close to $\theta$. A difference between $\tau_{0}$ and $\theta$ is small
at small pulse amplitude $a{\cal E}\ll V$. For this reason, the ``time'' interval $(\tau_{0}-\theta)$ near $\tau_{0}$, when
${\cal E}(i\tau)$ is not small, weakly contributes to the action. During the short ``time'' $(\tau_{0}-\theta)$ the particle
energy reduces (because ${\cal E}<0$) down to $(E-|\delta E|)$, so that
\begin{equation}
\label{29}
\tau_{00}(E-|\delta E|)\simeq\theta
\end{equation}
and the particle motion at $0<\tau <\theta$ can be considered to be free
\begin{equation}
\label{30}
A=\frac{2}{\hbar}\int^{\theta}_{0}\left[\frac{m}{2}\left(\frac{\partial x}{\partial\tau}\right)^{2}+V(x)-E\right]
\end{equation}
Adding and subtracting the energy transfer $|\delta E|$ in the right-hand side of Eq.~(\ref{30}) and using the equation 
(\ref{29}), one can arrive at
\begin{equation}
\label{31}
A=A_{0}(E-|\delta E|)-\frac{2\theta}{\hbar}|\delta E|
\end{equation}
With account of Eqs.~(\ref{28}) and (\ref{29}), it is obvious, that Eqs.~(\ref{12}) and (\ref{31}) give the same result.
\subsection{Why a small pulse may enhance tunneling} 
It is already been shown in a simple way in \ref{sec:ph-as}, that the effective nonstationary field in tunneling processes 
enhances compared to other ones. This enhancement is clearly seen in the method of trajectories in imaginary time where
an influence of a pulse becomes substantial under the approximate condition ${\cal E}(i\tau)\sim{\cal E}_{0}$. Here 
${\cal E}_{0}\sim V/a$ is the field of a static barrier. In the case of the pulse (\ref{4}), 
${\cal E}(i\tau)\sim{\cal E}\theta/(\tau -\theta)$ is enhanced compared to ${\cal E}$ since one can choose $\tau$ very close to 
$\theta$. But this enhancement is not unlimited due to the semiclassical condition of slow varying in time 
$(\tau -\theta)\ll\hbar/V$. Thus, the condition on the pulse amplitude coincides with (\ref{7}) or it can be written as
\begin{equation}
\label{31a}
\frac{1}{A_{0}}<\frac{{\cal E}}{{\cal E}_{0}}
\end{equation}
Since $A_{0}\sim V\theta/\hbar$ is big, tunneling can be strongly influenced by a pulse amplitude ${\cal E}$ which is less then
the static field of the barrier. This statement is true for more general forms of non-stationary fields. For example, the
monochromatic ${\cal E}\cos\Omega t$ and the Gaussian ${\cal E}\exp(-\Omega^{2}t^{2})$ pulses become exponentially big in imaginary
time and their stimulation of tunneling occurs at small amplitudes ${\cal E}<{\cal E}_{0}$ as well.
\subsection{Euclidean resonance in real time}
The metric in relativity $x^{2}+y^{2}+z^{2}-c^{2}t^{2}$ is Euclidean in imaginary time $t=i\tau$, since the all coefficient
become positive, and the action (\ref{26}) or (\ref{30}) is called Euclidean action. Extending this analogy, one can name 
the present phenomenon Euclidean resonance. 

The above approach is valid when $\exp(-A)\ll 1$, but upon reduction of $A$ this condition may be violated. When $\exp(-A)$ 
becomes no small one should account further contributions $\exp(-2A)$, $\exp(-3A)$, etc. This is equivalent to an account of 
multi-instanton contributions. 

\begin{figure}[p]
\begin{center}
\vspace{1.5cm}
\leavevmode
\epsfbox{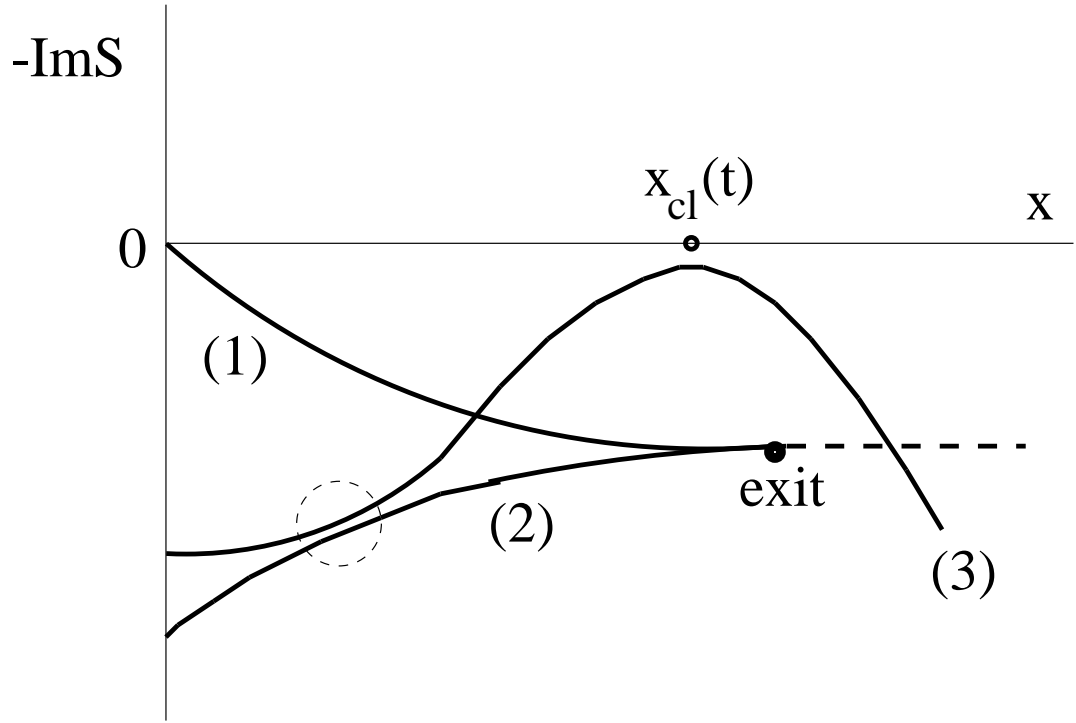}
\vspace{2cm}
\caption{The branches of the wave function $\psi\sim\exp(-{\rm Im}S)$. In the absence of a nonstationary field there are the
dominant branch $(1)$ and the sub-dominant one $(2)$ which merge at the point ``exit'' and convert into the outgoing wave, 
denoted by the dashed line. With a nonstationary field, at $t=0$ the branch $(3)$ is generated from the circled region at 
the sub-dominant branch $(2)$ where the semiclassical condition is violated at $t=0$. The branch $(3)$ is formed during a 
short (non-semiclassical) time and then it moves semiclassically, keeping its maximum at the classical trajectory point 
$x_{cl}(t)$. This maximum value of the branch $(3)$ decreases slowly in time due to quantum effects of smearing.}
\label{fig6}
\end{center}
\end{figure}

For the particular barrier $V(x)=V-x{\cal E}_{0}$ in presence of a non-stationary field one can build up a bridge between 
trajectories in imaginary time and the solution of Schr\"{o}dinger equation in real time. One can find this solution in 
the form $\psi(x,t)=a(x.t)\exp[iS(x,t)/\hbar]$, where $S(x,t)$ is the classical action and one can obtain any correction in $\hbar$ 
in the pre-exponent $a(x,t)$. The result for the imaginary part of the action is shown schematically in Fig.~\ref{fig6}. 
Without a pulse, the solution to the left of the point ``exit'', $\psi=c_{1}\psi_{1}+c_{2}\psi_{2}$ consists of the dominant, 
$\psi_{1}\sim\exp(iS_{1}/\hbar)$, and the sub-dominant, $\psi_{2}\sim\exp(iS_{2}/\hbar)$, solutions. 
$\psi_{2}(x=0)\sim\psi_{1}(x=0)\exp(-A_{0})$ is exponentially small, at the point ``exit'' $\psi_{1}\sim\psi_{2}\sim\exp(-A_{0}/2)$, and to
the right of ``exit'' there is only an outgoing wave of the amplitude $\exp(-A_{0}/2)$. This is a picture of decay of the 
metastable state, localized near $x=0$, through a static barrier.

When ${\cal E}(t)$ is not zero, the third solution, $\psi_{3}\sim\exp(iS_{3}/\hbar)$, appears which is shown by the curve (3) in
Fig.~\ref{fig6}. The maximum of this solution 
\begin{equation}
\label{32}
\psi_{3}\left[x_{cl}(t),t\right]\sim c_{3}\exp(-A/2)
\end{equation}
($A$ is defined by Eq.~(\ref{26})) is reached at the classical trajectory $x_{cl}(t)$ in real time, when 
$x_{cl}(\pm\infty)=\infty$  and $x_{cl}(0)$ is the minimum value. We omit not strong effects of quantum smearing of the wave 
packet. At a fixed moment of time $t<0$ one can find $\psi_{3}(x,t)$ for all $x$ using the generalized semiclassical
approach which nowhere breaks down upon sweeping over all $x$. In this case, $\psi_{3}$ is an independent solution and one
should put $c_{3}=0$ at $t<0$ since there is no incoming wave. Close to the moment $t=0$ the solution $\psi_{2}$ and  $\psi_{3}$
get a tendency to merge at some point, circled by the dashed curve in Fig.~\ref{fig6}. In this region 
$\partial^{2}S/\partial x^{2}$, calculated semiclassically, becomes big and the semiclassical approximation breaks down in a 
vicinity of the circled point. This means, that within the short (non-semiclassical) time interval $t\sim\hbar/V$ the 
solution $\psi_{3}$ is formed. Then, at $t>0$, the semiclassical solution recovers at all $x$ again, but now $c_{3}$ is not
zero, which relates to an outgoing particle. From a semiclassical point of view, there is a jump (since a semiclassical 
approximation does not resolve a short time) from zero to one of the coefficient at the third solution at the moment $t=0$.

Despite of that the above solution is obtained analytically for the barrier $V(x)=V-x{\cal E}_{0}$, the presented scenario of 
stimulation of tunneling by a nonstationary field holds qualitatively for a general semiclassical barrier. The jump of 
the coefficient occurs both in photon-assisted tunneling and in Euclidean resonance.
\section{EUCLIDEAN RESONANCE AND NUCLEAR REACTIONS}
\label{sec:nucl}
In \ref{sec:nucl} we apply the developed ideas of stimulation of tunneling to nuclear reactions where tunneling through a
Coulomb barrier is a substantial part of the process. The role of nonstationary field is played now by a charged incident
particle.
\subsection{Alpha decay of nuclei}
\begin{figure}[p]
\begin{center}
\vspace{1.5cm}
\leavevmode
\epsfxsize=\hsize
\epsfxsize=9cm
\epsfbox{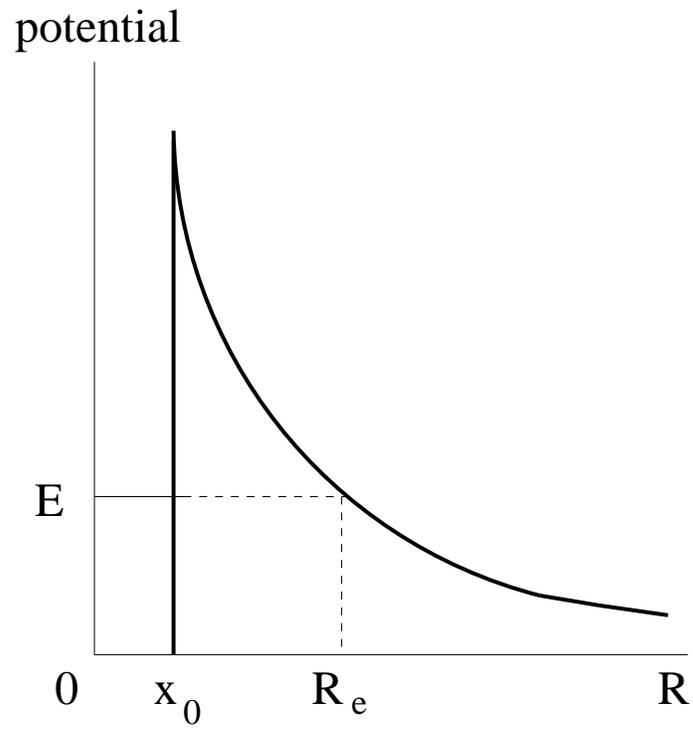}
\vspace{1cm}
\caption{The potential barrier which is passed by the alpha particle. The nuclear Coulomb field is cut off at the nuclear 
radius $x_{0}$. $R_{e}$ is the classical exit point.}
\label{fig7}
\end{center}
\end{figure}

According to Gamov \cite{WEISSKOPF,BOHR}, alpha decay of nuclei, is described by tunneling of alpha particle through the 
Coulomb barrier. The potential energy, as a function of the distance $R$ between alpha particle and the nucleus, is shown 
in Fig.~\ref{fig7}, where the Coulomb tail $\alpha_{M}/R$ ($\alpha_{M}=2(z_{0}-2)e^{2}$) sharply drops at the nuclear size 
$x_{0}$. For the alpha decay 
\begin{equation}
\label{33}
^{235}_{92}{\rm U}\rightarrow\hspace{0.1cm}^{231}_{90}{\rm Th}+\alpha
\end{equation}
$z_{0}=92$, $E=4.678$ Mev, and, according to the liquid drop model, 
\begin{equation}
\label{34}
x_{0}=1.2\left[(231)^{1/3}+4^{1/3}\right]\times 10^{-13}{\rm cm}\simeq 0.92\times 10^{-12}{\rm cm}
\end{equation}
The WKB tunneling rate is
\begin{equation}
\label{35}
W\sim\exp\left[-A_{M}(E,L_{M})\right]
\end{equation}
where the action in units of Planck's constant is
\begin{equation}
\label{36}
A_{M}(E,L_{M})=\frac{\sqrt{8M}}{\hbar}\int^{R_{e}}_{x_{0}}dR\sqrt{\frac{\alpha_{M}}{R}+\frac{L^{2}_{M}}{2MR^{2}}-E}
\end{equation}
$M$ is the mass of alpha particle and the classical exit point $R_{e}$ is determined by zero of the square root. 
$L_{M}\gg\hbar$ is the angular momentum. Further we consider parameters under the condition
\begin{equation}
\label{37}
\frac{L^{2}_{M}}{2Mx^{2}_{0}}\sim\frac{\alpha_{M}}{x_{0}}\gg E
\end{equation}
The action can be written in the form
\begin{equation}
\label{38}
A_{M}(E,L_{M})=\frac{\pi\alpha_{M}}{\hbar}\sqrt{\frac{2M}{E}}\left[1-\frac{4}{\pi}\sqrt{\frac{Ex_{0}}{\alpha_{M}}}f(p)\right]
\end{equation}
where
\begin{equation}
\label{39}
f(p)=\sqrt{1+p}-\frac{\sqrt p}{2}\ln\left[1+2p+2\sqrt{p(1+p)}\right]\hspace{0.1cm};\hspace{1cm}
p=\frac{L^{2}_{M}/2Mx^{2}_{0}}{\alpha_{M}/x_{0}}
\end{equation}
$p$ is the ratio of the centrifugal energy and the Coulomb one at the nucleus radius. The imaginary time of motion under 
the barrier 
\begin{equation}
\label{40}
\tau_{M}(E)=-\frac{1}{2}\hspace{0.1cm}\frac{\partial A_{M}(E,L_{M})}{\partial E}\simeq\frac{\pi\alpha_{M}}{2\hbar E}\sqrt{\frac{M}{2E}}
\end{equation}
weakly depends on the angular momentum $L_{M}$ under conditions (\ref{37}). 

According to the semiclassical applicability, the correction to $A_{M}$ due to the second term in Eq.~(\ref{38}) should be 
much bigger then one, that is
\begin{equation}
\label{41}
\sqrt{\frac{\hbar^{2}}{32M\alpha_{M}}}\ll\sqrt{x_{0}}
\end{equation}
This coincides with the conventional WKB condition in Coulomb field. In this problem of alpha decay the condition 
(\ref{41}) reads as $0.01\ll 1$. Since the parameter $p\sim 1$, the angular momentum is big
\begin{equation}
\label{42}
\frac{L_{M}}{\hbar}\sim\sqrt{\frac{M\alpha_{M}}{\hbar^{2}x_{0}}}\gg 1
\end{equation}
As one can see, semiclassical conditions are fulfilled well for alpha decay.
\subsection{An incident proton}
\begin{figure}[p]
\begin{center}
\vspace{6cm}
\leavevmode
\epsfbox{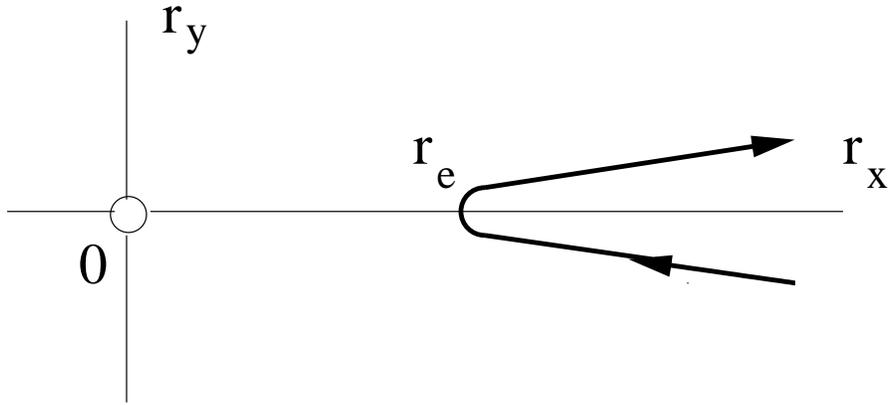}
\vspace{1cm}
\caption{The classical trajectory of the proton moving in the nucleus Coulomb field. $r_{e}$ is the classical turning 
point.}
\label{fig8}
\end{center}
\end{figure}

What happens to alpha decay when a charged particle (proton, for example) is stopped by the Coulomb field of the nucleus 
which is ready to emit alpha particle? The classical motion of the proton in the Coulomb field of uranium nucleus is 
described by the equation
\begin{equation}
\label{43}
t=\sqrt{\frac{m}{2}}\int dr\left(\varepsilon -\frac{L^{2}_{m}}{2mr^{2}}-\frac{\alpha_{m}}{r}\right)^{-1/2}
\end{equation}
where $m$ is proton mass, $\alpha_{m}=z_{0}e^{2}$, $L_{m}$ is the proton angular momentum, and $\varepsilon$ is the proton 
energy. The classical trajectory is shown in Fig.~\ref{fig8}, where the shortest distance $r_{e}$ between the proton and 
the nucleus (the classical turning point) is given by the zero of the square root in Eq.~(\ref{43}). At $r<r_{e}$ the time 
$t$ in Eq.~(\ref{43}) becomes imaginary and, starting at the point $r_{e}$, the proton reaches the nucleus (the dashed line 
in Fig.~\ref{fig8}) at the ``moment'' $t=i\tau_{m}$.  Analogously to Eq.~(\ref{40}), the expression for $\tau_{m}$ is
\begin{equation}
\label{44}
\tau_{m}(\varepsilon)=\frac{\pi\alpha_{m}}{2\hbar\varepsilon}\sqrt{\frac{m}{2\varepsilon}}
\end{equation}
\subsection{Alpha particle meets proton}
When the uranium nucleus emits the alpha particle, the additional interaction energy 
\begin{equation}
\label{45}
V_{int}=\frac{\alpha_{int}}{|\vec{R}(i\tau)-\vec{r}(i\tau)|}
\end{equation}
where $\alpha_{int}=2e^{2}$, results in a connection of motions of the alpha particle and the proton. Now there is a
cooperative motion of two particles in imaginary time which starts at $\tau =0$, with zero radial velocities, and terminates
at the nucleus at the certain under-barrier time $\tau_{0}$. The total energy $(E+\varepsilon)$ of two particles conserves. 
$\varepsilon$ is the energy of the incident proton and $E$ is the energy of the alpha particle close to the nucleus at
$\tau =\tau_{0}$. The interaction energy is always small excepting a narrow vicinity of the moment $\tau =\tau_{0}$ when
two particles are close to the nucleus where $|\vec{R}-\vec{r}|\sim x_{0}$. In the small vicinity of $\tau_{0}$ the interaction
redistributes energies, so that the alpha particle leaves the interaction region with the energy $E-|\delta E|$ and the 
proton energy becomes $\varepsilon +|\delta E|$. The major part of the interval $(\tau_{0},0)$, excepting a small vicinity of 
$\tau_{0}$, the particles move independently with the redistributed energies, reach the point $\tau =0$, and go to infinity
in real time having the same energies $E-|\delta E|$ and $\varepsilon +|\delta E|$. The interaction (\ref{45}) contributes weakly
to the action since it works during a short time. 

\begin{figure}[p]
\begin{center}
\vspace{1.5cm}
\leavevmode
\epsfxsize=\hsize
\epsfxsize=9cm
\epsfbox{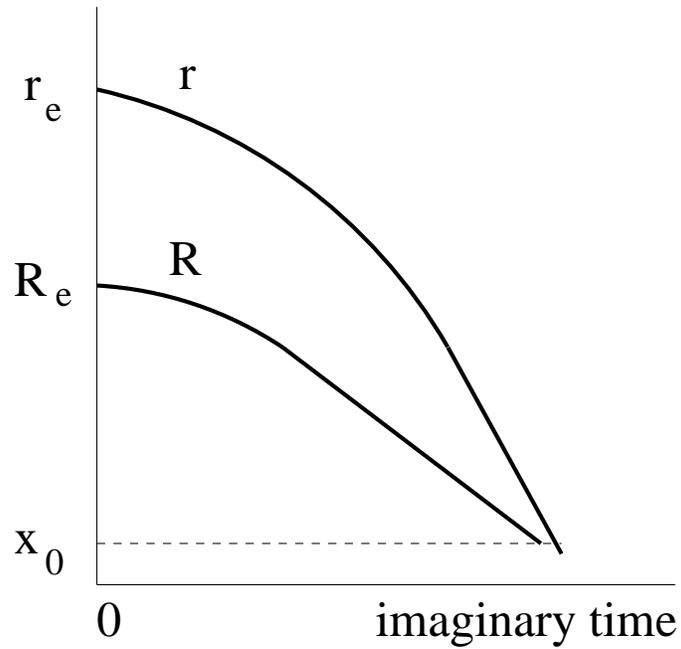}
\vspace{1cm}
\caption{The classical trajectories in imaginary time of the proton $r$ and of the alpha particle $R$. The both 
particles occur at the nucleus position $x_{0}$ at the same ``moment'' of imaginary time.}
\label{fig9}
\end{center}
\end{figure}

The both trajectories are shown in Fig.~\ref{fig9}, where the exit points are $R_{e}=\alpha_{M}/(E-|\delta E|)$ and
$r_{e}=\alpha_{m}(\varepsilon +|\delta E|)$. As $x_{0}$ is small, the both curves are about to merge at $\tau =\tau_{0}$, otherwise
the interaction at this region is not effective. This requires the condition
\begin{equation}
\label{46}
\tau_{M}(E-|\delta E|)=\tau_{m}(\varepsilon +|\delta E|)\simeq\tau_{0}
\end{equation}
With Eqs.~(\ref{40}) and (\ref{44}) the condition (\ref{46}) reads
\begin{equation}
\label{47}
\frac{\varepsilon +|\delta E|}{E-|\delta E|}=\left(\frac{m\alpha^{2}_{m}}{M\alpha^{2}_{M}}\right)^{1/3}
\end{equation}
\subsection{Cooperative motion of alpha particle and proton}
The cooperative motion of the alpha particle and the proton relates to the Euclidean action
\begin{equation}
\label{48}
\tilde{A}=\frac{2}{\hbar}\int^{\tau_{0}}_{0}d\tau
\left[\frac{M}{2}\left(\frac{\partial\vec{R}}{\partial\tau}\right)^{2}+\frac{\alpha_{M}}{R}+
\frac{m}{2}\left(\frac{\partial\vec{r}}{\partial\tau}\right)^{2}+\frac{\alpha_{m}}{r}+\frac{\alpha_{int}}{|\vec{R}-
\vec{r}|}-E-\varepsilon\right]
\end{equation}
The classical trajectory satisfies Newton's equation resulting from a minimization of $\tilde{A}$ with the initial 
conditions
\begin{equation}
\label{49}
\frac{\partial R_{x}}{\partial\tau}\bigg |_{0}=\frac{\partial r_{x}}{\partial\tau}\bigg |_{0}=0\hspace{0.1cm};\hspace{0.5cm}
R_{y}(0)=r_{y}(0)=0\hspace{0.1cm};\hspace{0.5cm}MR_{x}\hspace{0.1cm}\frac{\partial R_{y}}{\partial\tau}\bigg |_{0}=L_{M}\hspace{0.1cm};
\hspace{0.5cm}mr_{x}\hspace{0.1cm}\frac{\partial r_{y}}{\partial\tau}\bigg |_{0}=L_{m}
\end{equation}
where $L_{M}$ and $L_{m}$ are the angular momenta of particles which conserve for all $\tau$ excepting a close vicinity of
$\tau_{0}$. The another condition is
\begin{equation}
\label{50}
R(\tau_{0})=r(\tau_{0})=x_{0}
\end{equation}
and $\tau_{0}$ relates to the energy of alpha particle 
\begin{equation}
\label{51}
E=\left[-\frac{M}{2}\left(\frac{\partial\vec{R}}{\partial\tau}\right)^{2}+\frac{\alpha_{M}}{R} \right]_{\tau_{0}}
\end{equation}
In Eqs.~(\ref{48})-(\ref{51}) the vectors are defined as $\vec{R}=\{R_{x},iR_{y}\}$ and $\vec{r}=\{r_{x},ir_{y}\}$ since in imaginary 
time $y$-components are imaginary. This can be seen in the proton motion in the Coulomb field of the nucleus (with no 
alpha particle) under the condition analogous to (\ref{37}) for proton
\begin{equation}
\label{52}
r_{x}=\left(r+\frac{L^{2}_{m}}{m\alpha_{m}}\right)\left(1+\frac{2\varepsilon L^{2}_{m}}{m\alpha_{m}}\right)^{-1/2};\hspace{0.5cm}
r_{y}=\sqrt{\frac{L^{2}_{m}}{2m\varepsilon}+\frac{r\alpha_{m}}{\varepsilon}-r^{2}}
\left(1+\frac{m\alpha_{m}}{2\varepsilon L^{2}_{m}}\right)^{-1/2}
\end{equation}
The dependence $r(i\tau)$ is given by Eq.~(\ref{43}) with $t=i\tau$. As follows from Eq.~(\ref{52}), $r^{2}_{x}-r^{2}_{y}=r^{2}$.
Fall of the proton to the nucleus means that $r(i\tau)=0$, but $r_{x}(i\tau)$ and $r_{x}(i\tau)$ separately are not zero.

In contrast to the big energy transfer $|\delta E|$, the interaction (\ref{45}) results in a very small transfer of angular 
momentum $\delta L_{m}$ between alpha particle and proton. On the basis of the classical relation 
\begin{equation}
\label{53}
\frac{\partial\delta L_{m}}{\partial\tau}=\frac{\alpha_{int}}{|\vec{R}-\vec r|^{3}}(r_{x}R_{y}-r_{y}R_{x})
\end{equation}
with the estimate $R_{x}\sim r_{x}\sim |\vec{R}-\vec{r}|\sim \alpha_{m}/\varepsilon$ and Eq.~(\ref{52}) for the $y$-component, one can 
easily deduce, that $\delta L_{m}/L_{m}\sim\alpha_{int}/\alpha{m}\ll 1$. As shown below, the energy transfer $|\delta E|$ is not small
since it is determined by by small $|\vec{R}-\vec{r}|\sim x_{0}$. The proton can change its angular momentum solely by 
interaction with the nucleus. We consider only spherically symmetric nuclei and the proton angular momentum $L_{m}$ 
conserves. The angular momentum $L_{M}$ of the alpha particle is determined by the interaction with the nucleus and 
conserves for all $\tau$. 

Now it is possible to calculate $\tilde{A}$ in Eq.~(\ref{48}). A contribution to $\tilde{A}$ from the interaction part $V_{int}$
is small, almost all ``time'' $\tau$ the alpha particle conserves its energy $E-|\delta E|$, and the same is true for the 
proton with the energy $\varepsilon-|\delta E|$. For this reason, one can write $\tilde{A}$ in Eq.~(\ref{48}) as a sum of two
actions of free particles 
\begin{equation}
\label{54}
\tilde{A}=A_{M}(E-|\delta E|, L_{M})+A_{m}(\varepsilon+|\delta E|, L_{m})
\end{equation}
where $A_{m}$ is determined by Eq.~(\ref{38}) with the substitution $M\rightarrow m$. $|\delta E|$ obeys Eq.~(\ref{47}), $L_{m}$ is
given by the angular momentum of the incident proton, and $L_{M}$ is obtained by the alpha particle from the nucleus. We 
do not distinguish in $\alpha_{m}=z_{0}e^{2}$ between $z_{0}$ and $(z_{0}-2)$ as the correction is of the same order as the small
interaction. 

Since the energy transfer $|\delta E|$ is determined by small $|\vec{r}-\vec{r}|\sim x_{0}$, it strongly depends on angular momenta
$L_{M}$ and $L_{m}$ which set a smallest distance between particles. $L_{M}$ and $L_{m}$ should be chosen to get the energy
transfer $|\delta E|$ obeyed Eq.~(\ref{47}). We discuss this below.
\subsection{Phase connection}
\begin{figure}[p]
\begin{center}
\vspace{5cm}
\leavevmode
\epsfxsize=\hsize
\epsfxsize=15cm
\epsfbox{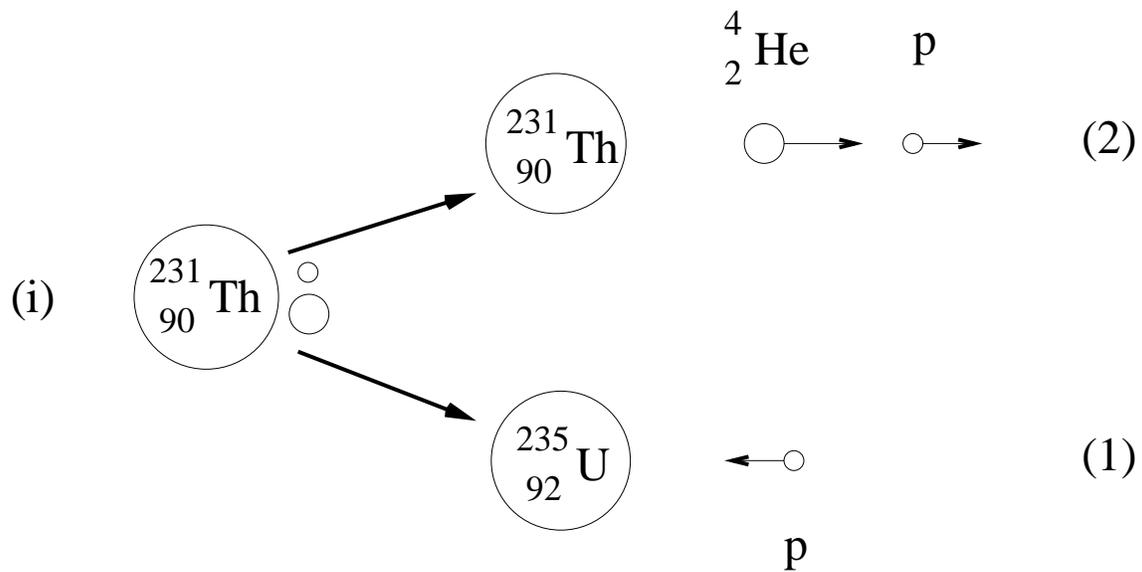}
\vspace{1cm}
\caption{$(1)$ is the initial state of the reaction which includes the uranium nucleus and the incident proton. $(2)$ is 
the final state, including the outgoing alpha particle and the proton. The state $(i)$ consists of the thorium nucleus 
with the alpha particle and the proton close to it, but away of nuclear forces. The state $(i)$ serves to connect the 
phases of the states $(1)$ and $(2)$.}
\label{fig10}
\end{center}
\end{figure}

The action $\tilde{A}$ relates to the phase connection
\begin{equation}
\label{55}
\psi_{2}\sim\psi_{i}\exp\left(-\frac{\tilde{A}}{2}\right)
\end{equation}
of two states, $(i)$ and (2), shown in Fig.~\ref{fig10}, where (2) is the physical final state of the reaction. The state 
$(i)$ consists of the nucleus $^{231}_{90}$Th with the alpha particle and proton close to it. (1) is the physical initial 
state with the incident proton. The phase connection between the states $(i)$ and (1) reads
\begin{equation}
\label{56}
\psi_{1}\sim\psi_{i}\exp\left[-\frac{1}{2}A_{m}(\varepsilon,L_{m})\right]
\end{equation}
The dominant contribution to the imaginary phases comes from motion in the Coulomb field. By means of Eqs.~(\ref{55}) and 
(\ref{56}) one can obtain the probability $W$ of the reaction
\begin{equation}
\label{57}
^{235}_{92}{\rm U}+p\rightarrow\hspace{0.1cm}^{231}_{90}{\rm Th}+\alpha +p
\end{equation}
in the form
\begin{equation}
\label{58}
W\sim\bigg |\frac{\psi_{2}}{\psi_{1}}\bigg |^{2}\sim\exp{-A}
\end{equation}
where
\begin{equation}
\label{59}
A=A_{M}(E-|\delta E|, L_{M})+A_{m}(\varepsilon+|\delta E|, L_{m})-A_{m}(\varepsilon, L_{m})          
\end{equation}
This procedure of phase connection is same as in \ref{sec:eucl}.

One can use for the action (\ref{59}) the approximation (\ref{38}), which is used in the derivation of Eq.~(\ref{47}). In 
this approximation $A$ does not depend on $L_{m}$ due to the cancellation in the last two terms of Eq.~(\ref{59}). This 
means, that $L_{m}$ can vary to adjust the energy transfer $|\delta E|$ to one given by Eq.~(\ref{47}) with no impact on $A$. 
In this situation the minimum $A$ is reached at $L_{M}=0$. Finally, the action in Eq.~(\ref{59}) takes the form
\begin{equation}
\label{60}
A=\frac{\pi\alpha_{M}}{\hbar}\sqrt{\frac{2M}{E-|\delta E|}}-4\sqrt{\frac{x_{0}}{\hbar^{2}/2M\alpha_{M}}}+
\frac{\pi\alpha_{m}\sqrt{2m}}{\hbar}\left(\frac{1}{\sqrt{\varepsilon +|\delta E|}}-\frac{1}{\sqrt{\varepsilon}}\right)         
\end{equation}
where $|\delta E|$ satisfies the relation (\ref{47}).

\begin{figure}[p]
\begin{center}
\vspace{6cm}
\leavevmode
\epsfxsize=\hsize
\epsfbox{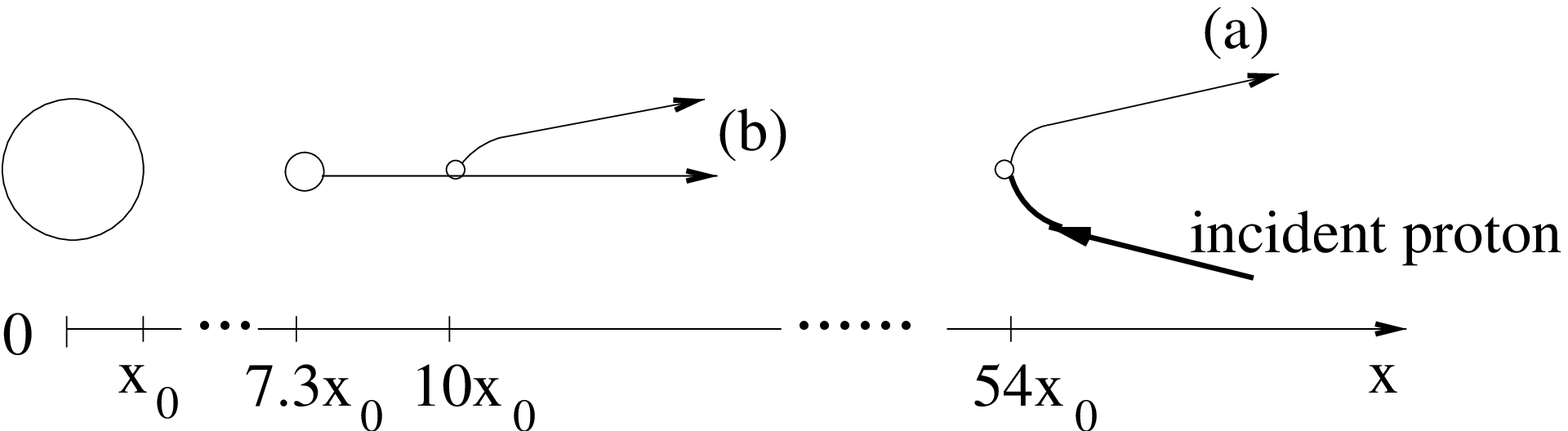}
\vspace{1cm}
\caption{The big circle is the nucleus. The incident proton (the smallest circle) moves along the thick curve and may 
reflect elastically with no stimulation of alpha decay $(a)$. The incident proton can initiate the alpha decay and then it
appears, together with the alpha particle (the small circle), in the channel $(b)$. The exit points at the $x$-axis are
given for the energy of the incident proton $0.25$\hspace{0.1cm}Mev.}
\label{fig11}
\end{center}
\end{figure}

\begin{figure}[p]
\begin{center}
\vspace{1.5cm}
\leavevmode
\epsfbox{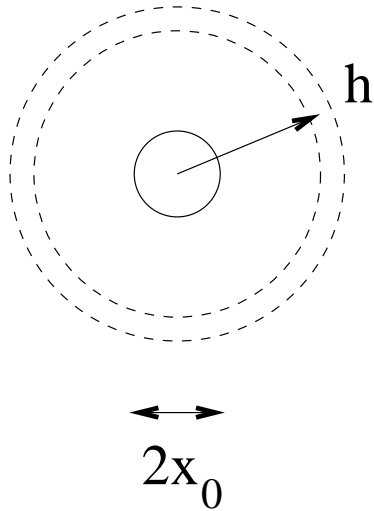}
\vspace{1cm}
\caption{$h$ is the impact parameter of the incident proton. The cross-section of the reaction is determined by the area
between two dashed circles, separated by the distance $\delta h$.}
\label{fig12}
\end{center}
\end{figure}

The physical trajectories of particles in the nuclear reaction (\ref{57}) are shown in Fig.~\ref{fig11}. The probability of
the channel $(a)$ in Fig.~\ref{fig11} is almost 100$\%$ and the probability of the cannel $(b)$, which is the reaction
(\ref{57}), is of the order of $\exp(-A)\ll 1$. In Fig.~\ref{fig12} the classical positions of particles are shown at the
moment of time when the incident proton reaches its minimum distance to the nucleus. 

The angular momentum $L_{m}$ of the proton should be exactly of the value to provide the energy transfer (\ref{47}). 
Otherwise, the proton and the alpha particle do not meet at the nucleus position at the coincident ``times''
$\tau_{M}=\tau_{m}$, their interaction would be small and the action would be not a result of a cooperative motion of two
particles but simply conventional $A_{M}(E,L_{M})$.
\subsection{Required angular momentum of the incident proton}
Let us calculate the energy transfer $|\delta E|$ at zero angular momentum of the incident proton $L_{m}=0$ and then find 
the proton energy $\varepsilon$ which corresponds to this process. Since $L_{m}=0$, only $x$-components are involved which
are determined by the classical dynamical equations
\begin{equation}
\label{61}
M\hspace{0.1cm}\frac{\partial R_{x}}{\partial\tau^{2}}=-\frac{\alpha_{M}}{R^{2}_{x}}+\frac{\alpha_{int}}{(r_{x}-R_{x})^{2}}
\hspace{0.1cm};
\hspace{1cm}m\hspace{0.1cm}\frac{\partial r_{x}}{\partial\tau^{2}}=-\frac{\alpha_{m}}{r^{2}_{x}}-\frac{\alpha_{int}}{(r_{x}-R_{x})^{2}}
\end{equation}
close to $\tau_{0}$ ($\tau_{0}\simeq\tau_{M}\simeq\tau_{m}$) the solutions have the form
\begin{equation}
\label{62}
\frac{R_{x}(i\tau)}{R_{s}}=\frac{r_{x}(i\tau)}{r_{s}}=\left(\frac{\tau_{m}-\tau}{\tau_{m}}\right)^{2/3}
\end{equation}
where $R_{s}$ and $r_{s}$ are some constants. The energy $\delta E$, gained by the alpha particle, 
\begin{equation}
\label{63}
\delta E=\alpha_{int}\int^{\tau_{0}}_{0}\frac{d\tau}{(R_{x}-r_{x})^{2}}\hspace{0.1cm}\frac{\partial R_{x}}{\partial\tau}
\end{equation}     
diverges close to $\tau_{0}$ and should be cut off by the condition $R_{s}(1-\tau/\tau_{m})^{2/3}>x_{0}$. This gives
\begin{equation}
\label{64}
|\delta E|=\frac{\alpha_{int}}{x_{0}}\left(\frac{r_{s}}{R_{s}}-1\right)^{-2}
\end{equation} 
The ratio $R_{s}/r_{s}$, as one can see after a little algebra, satisfies the relation
\begin{equation}
\label{65}
\frac{M\alpha_{m}}{m\alpha_{M}}\left(\frac{R_{s}}{r_{s}}\right)^{3}
\left[\left(1-\frac{R_{s}}{r_{s}}\right)^{2}+\frac{\alpha_{int}}{\alpha_{m}}\right]=
\left(1-\frac{R_{s}}{r_{s}}\right)^{2}-\frac{\alpha_{int}}{\alpha_{M}}\left(\frac{R_{s}}{r_{s}}\right)^{2}
\end{equation}
Substituting parameters for the reaction (\ref{57}) $M/m=4$, $\alpha_{M}/\alpha_{m}=2$, and $\alpha_{int}/\alpha_{M}=1/90$,
one can obtain $R_{s}/r_{s}\simeq 0.715$ and the energy transfer $|\delta E|\simeq 1.89$\hspace{0.1cm}MeV. We use the 
estimate (\ref{34}) for the nuclear size. With this $|\delta E|$ and the alpha particle energy $E=4.678$\hspace{0.1cm}MeV,
the relation (\ref{47}) gives an unphysical (negative) value of $\varepsilon$. This means, that an incident proton with 
zero angular momentum transfers too big energy and the real process requires a finite angular momentum $L_{m}$ in order to
effectively increase the minimum distance $x_{0}$ in Eq.~(\ref{64}) to reduce the energy transfer. The minimum 
proton-nucleus distance increases when the proton centrifugal energy becomes of the order of the Coulomb one at the 
nucleus radius. This corresponds to the estimate (\ref{37}) for proton 
\begin{equation}
\label{65a}
\frac{L^{2}_{m}}{2mx^{2}_{0}}\sim\frac{\alpha_{m}}{x_{0}}
\end{equation}
and a typical angular momentum of the incident proton should be $L_{m}\sim 10\hbar$.
\subsection{Euclidean resonance}
If to insert the energy transfer $|\delta E|$ from Eq.~(\ref{47}) into Eq.~(\ref{60}), one can obtain
\begin{equation}
\label{66}
A=\frac{\pi\alpha_{M}}{\hbar}\sqrt{\frac{2M}{E+\varepsilon}}\left[1+\left(\frac{m\alpha^{2}_{m}}{M\alpha^{2}_{M}}\right)^{1/3}\right]^{3/2}-
\frac{4}{\hbar}\sqrt{2Mx_{0}\alpha_{M}}-\frac{\pi\alpha_{m}}{\hbar}\sqrt{\frac{2m}{\varepsilon}}
\end{equation}

At $\varepsilon =\varepsilon_{max}$, where
\begin{equation}
\label{67}
\varepsilon_{max}=E\left(\frac{m\alpha^{2}_{m}}{M\alpha^{2}_{M}}\right)^{1/3}\simeq 1.85\hspace{0.1cm}{\rm MeV}
\end{equation}
($E=4.678$\hspace{0.1cm}MeV) the energy transfer $|\delta E|=0$, the parameter $L^{2}_{m}/2mx_{0}\alpha_{m}$ has its maximum 
value, and the action (\ref{66}) matches the conventional one (\ref{38}) resulting in the tunneling probability
\begin{equation}
\label{68}
W\sim\exp\left[-A_{M}(4.678\hspace{0.1cm}{\rm MeV},0)\right]\simeq e^{-80.75}\simeq 10^{-35}
\end{equation}
The result (\ref{68}) reasonably describes experimental data if to multiply (\ref{68}) by the nuclear attempt frequency
$10^{21}\hspace{0.1cm}{\rm s}^{-1}$. 

At $\varepsilon <\varepsilon_{max}$ the energy transfer $|\delta E|$ becomes finite, $L_{m}$ decreases, and the action (\ref{66}) 
reduces compared to $A_{M}(E,0)$. Upon reduction of $\varepsilon$, the action (\ref{66}) turns to zero at the certain proton 
energy $\varepsilon_{R}$, which relates to Euclidean resonance. For the reaction (\ref{57}) $\varepsilon_{R}=0.25$\hspace{0.1cm}MeV 
and the accompanied energy transfer is $|\delta E|=1.15$\hspace{0.1cm}MeV. In other words, when in the reaction (\ref{57}) the 
energy of the incident proton is $\varepsilon =0.25$\hspace{0.1cm}MeV, it converts into the proton with the energy 
$\varepsilon +|\delta E| =1.40$\hspace{0.1cm}MeV and the energy of the emitted alpha particle (instead of 
$E=4.678$\hspace{0.1cm}MeV) becomes $E-|\delta E| =3.53$\hspace{0.1cm}MeV.

The cross-section of the reaction (\ref{57}) is not exponentially small at $\varepsilon =\varepsilon_{R}$, it has a sharp 
peak at this proton energy, and is determined by the angular momentum $L_{m}$ which provides the energy transfer 
$|\delta E|=1.15$\hspace{0.1cm}MeV. A rough estimate of this angular momentum (\ref{65a}) results in the impact parameter
$h\sim 8x_{0}$, according to the classical relation $L_{m}=h\sqrt{2m\varepsilon}$. A geometrical estimate of the cross-section at 
$\varepsilon =\varepsilon_{max}$ is reduced to the ring area $\sigma\simeq 2\pi h\delta h$ of the width $\delta h$ near the circle of 
the radius $h$, as shown in Fig.~\ref{fig12}. Since $h$ is some optimum value, related to an extreme action, $\delta h$ can 
be estimated as $\delta h/h\sim 1/\sqrt{A_{M}}$. By means of the relation (\ref{68}), the geometrical estimate of the 
cross-section produces 15 nuclear impact areas $\sigma\sim 15(\pi x^{2}_{0})$. 

At $\varepsilon =\varepsilon_{max}$ the exit point of alpha particle is $7.3x_{0}$ and one of the proton is $10x_{0}$. 
The incident proton is stopped by the nuclear Coulomb field at $54x_{0}$, as shown in Fig.~\ref{fig11}. The interaction of
alpha particle with a moving proton is analogous to its interaction with some non-stationary field, which results in
Euclidean resonance. As in Fig.~\ref{fig6}, the exponentially small part of alpha particle wave function merges at 
some moment of time the growing part of the wave function. Trajectories in imaginary time is only a convenient language to
describe the effect. In real (physical) time alpha particle does not approach the nucleus and interacts with it solely
via the Coulomb field. For example, this practically excludes spin interaction between them.

The above calculations hold for spherical nuclei. Alpha emitters may be not spherical, but real parameters of
nonsphericality unlikely essentially violate the estimates. 
\section{CONCLUSION}
When a proton approaches a nucleus, which is an alpha emitter, it creates a nonstationary Coulomb interaction with the 
tunneling alpha particle. At the certain proton energy there are conditions for Euclidean resonance and the Coulomb 
barrier becomes transparent for the  passage of the alpha particle. Normally, $^{235}_{92}$U emits alpha particle 
of the energy $4.678$\hspace{0.1cm}MeV. When the energy of the incident proton is close  to its resonant value 
$0.25$\hspace{0.1cm}MeV, it reflects with the energy $1.40$\hspace{0.1cm}MeV and simultaneously the alpha particle is 
emitted with the energy $3.53$\hspace{0.1cm}MeV. Beams of $0.2$\hspace{0.1cm}MeV-scale protons are ``cheap'' since they can 
be technically created in a relatively easy way. For this reason, low energy protons can be used for practical 
applications, for example, in disactivation of the nuclear waste.   

The analytical calculation of Euclidean resonance on the semiclassical basis is given in Ref.~\cite{IVLEV2}. In principal,
one can solve numerically the initial Schr\"{o}dinger equation for decay of a metastable state in presence of 
nonstationary field using the proper algorithm and accounting the boundary conditions \cite{GUDKOV}. However, there is a
serious problem in such numerical calculations. As one can see from Fig.~\ref{fig6}, the branch (3), related to Euclidean 
resonance, starts to form from the exponentially
small branch (2) which is of the order of $\exp(-A_{\rm WKB})$ at the well position. In order to get a numerical 
calculation accounted this effect, one should choose very short steps $\Delta t$ in time (the number of steps is 
proportional to $(\Delta t)^{-1}$) and $\Delta x$ in coordinate (the number of steps is proportional to $(\Delta x)^{-1})$. 
They have to satisfy the condition $\Delta t\sim(\Delta x)^{2}<\exp(-A_{\rm WKB})$, otherwise the effect would be missed. 
Suppose the numerical computation with $10^{3}$ steps in time and $10^{3}$ steps in coordinate requires one second. One 
can easily estimate the total computation time as $\exp(1.5A_{\rm WKB})\times 10^{-10}$~days. Since for alpha decay 
$A_{\rm WKB}\simeq 80$, the total computation time should be $10^{42}$~days. For this reason, numerical computation should
start not with the initial Schr\"{o}dinger equation but with some semiclassical approach.
\section{ACKNOWLEDGMENT}
I am grateful to V. Gudkov for very valuable discussions. I also appreciate discussions with R. Prozorov, M. Kirchbach,
J. Engelfried, J. M. Knight, and M. N. Kunchur.

\newpage

\end{document}